\documentclass[12pt,,letterpaper]{JHEP3}
\usepackage{graphics}
\usepackage{epsfig}

\title{A covariant entropy bound conjecture on the dynamical horizon
\footnote{Honorable Mention Award Received from Gravity Research Foundation for 2008 Essay Competition.}
}

\author{Song He\\
School of Physics, Peking
University, Beijing, 100871, China\\
 \email{hesong@pku.edu.cn}}

\author{Hongbao Zhang\\
Perimeter
Institute for Theoretical Physics, Waterloo, Ontario, N2L 2Y5,
Canada
\\Department of Astronomy, Beijing Normal University, Beijing,
100875, China\\
\email{hzhang@perimeterinstitute.ca}}

\abstract{As a compelling
pattern for the holographic principle, our covariant entropy bound
conjecture is proposed for more general dynamical horizons. Then we
apply our conjecture to $\Lambda$CDM cosmological models, where we
find it imposes a novel upper bound $10^{-90}$ on the
cosmological constant for our own universe by taking into account
the dominant entropy contribution from super-massive black holes,
which thus provides an alternative macroscopic perspective to
understand the longstanding cosmological constant problem. As an
intriguing implication of this conjecture, we also discuss the
possible profound relation between the present cosmological
constant, the origin of mass, and the anthropic principle.}

\begin{document}
\section{Introduction and motivation}
The generalized second law of thermodynamics was initially put forth
for a system including black holes by
Bekenstein\cite{Bekenstein1,Bekenstein2,Bekenstein3}. It states that
the sum of one quarter of the area of the black hole event horizon
plus the entropy of ordinary matter outside never decreases with
time in all processes. Especially, for the formation or absorption
of black holes, the generalized second law of thermodynamics can
also be equivalently formulated as a covariant entropy bound.
Namely, the entropy flux $S$ through the event horizon between its
two-dimensional space-like surfaces of area $A_e$ and $A_e'$ must
satisfy
\begin{equation}\label{eh}
S\leq\frac{A_e'-A_e}{4},
\end{equation}
where $A_e'\geq A_e$ is assumed, and Planck units are used, i.e.,
$c=G=\hbar=k=1$.

However, due to the global and teleological deficit of the event
horizon, the notion of the black hole dynamical horizon has recently
been developed quasi-locally to model growing black holes and its
properties have also been extensively investigated, where, in
particular, the first and second laws of black hole mechanics was
generalized to the dynamical horizon\cite{AK1,AK2,AK3}. Along this
line further, we have proposed a covariant entropy bound formulation
of an analogous generalized second law of thermodynamics on the
black hole dynamical horizon and its generalization to cosmological
dynamical horizons in FRW universes has also been
conjectured\cite{HZ1,HZ2}. Moreover, its validity has been confirmed
in both Vaidya black holes and FRW universes full of matter with a
fixed state of equation $w=\frac{p}{\rho}$, regardless of the
spatial geometry\cite{HZ1,HZ2}. All of these results suggest that
our proposal, viewed as a covariant entropy bound conjecture on
dynamical horizons, may be a universal law and there may be some
deep reasons for its validity. In fact, the conjecture is motivated
partly by Bousso's covariant entropy bound
conjecture\cite{Bousso1,Bousso2,Bousso3,Bousso4}, and its
strengthened form suggested by Flanagan, Marolf, and
Wald\cite{Flanagan}. These various entropy bound conjectures,
including ours, can also be interpreted as a statement of the so
called holographic principle, which is believed to be manifest in an
underlying quantum gravity\cite{Hooft,Susskind}.

Taking into account its success in many respects and justification
as a possible fundamental principle, we have quite recently applied
our covariant entropy bound conjecture to constrain those
cosmological models with a positive cosmological constant plus the
matter content satisfying the dominant energy condition\cite{HZ3}.
Especially, for $\Lambda$CDM cosmological models, it is found that
our conjecture implies a novel inequality as
\begin{equation}
s\sqrt{\Lambda}\leq 2\sqrt{3}\pi\rho^0_m,\label{novel}
\end{equation}
where $s$ represents the present entropy density, and $\rho^0_m$
denotes the energy density of dust today. This is a remarkable
result because it establishes a significant relation governing the
cosmological constant, present entropy density and dust energy
density.

In this essay, we further extend our covariant entropy bound
conjecture to more general dynamical horizons.  We then explore its
intriguing physical implications after reworking out the inequality
(\ref{novel}). Conclusion and discussion are presented in the end.

The signature of metric takes $(-,+,+,+)$. Notation and conventions
follow Ref.\cite{Wald}.
\section{A covariant entropy bound conjecture on
the dynamical horizon} We would first like to introduce the basic
definition of our dynamical horizon in a more general sense. Roughly
speaking, a dynamical horizon is just a hyper-surface which is
foliated by closed apparent horizons. A more detailed definition can
be presented as follows:

\emph{Definition}: A three-dimensional sub-manifold in a spacetime
($M, g_{ab}$) is said to be a dynamical horizon if it can be
foliated by a family of closed two-dimensional surfaces such that,
on each leaf, the expansion  $\theta_l$ of one future-directed null
normal $l^a$ vanishes while the expansion $\theta_n$ of the other
future-directed null normal $n^a$ is positive or negative, in
addition, the Lie derivatives of $\theta_l$ along $l^a$ and $n^a$ do
not vanish simultaneously, where, the normalization of $l^a$ and
$n^a$ is chosen such that $l^an_a=-2$, which implies the expansion
of the null geodesics normal can be given by
$\theta_l=h^{ab}\nabla_al_b$($\theta_n=h^{ab}\nabla_an_b$) with the
induced metric $h_{ab}=g_{ab}+\frac{1}{2}(l_an_b+n_al_b)$ on each
leaf.

\begin{table}\small
  \centering
  \begin{tabular}{|l|l|l|}
  \hline
 $\theta_l=0$ & $\theta_n>0$ & $\theta_n<0$ \\
 \hline
 $\pounds_n\theta_l>0,\pounds_l\theta_l<0$ & expanding FRW universes  &  time reversal \\
 (timelike)&   with $-1< w<\frac{1}{3}$&\\
 \cline{2-3}
 & time reversal & growing Vaidya-De sitter  \\
 && black holes\\
 \hline
  $\pounds_n\theta_l>0,\pounds_l\theta_l=0$ & expanding De sitter spacetime & time reversal \\
(null generated by $l^a$)&&\\
  \hline
   $\pounds_n\theta_l<0,\pounds_l\theta_l<0$& time reversal & growing Vaidya black holes \\
   \cline{2-3}
   (spacelike)&expanding FRW universes & time reversal\\
   &with $\frac{1}{3}<w\leq 1$&\\
   \hline
  $\pounds_n\theta_l<0,\pounds_l\theta_l=0$& time reversal & Schwarzchild black holes  \\
  (null generated by $l^a$)&&\\
  \hline
  $\pounds_n\theta_l=0,\pounds_l\theta_l<0$& expanding FRW universe  & time reversal\\

  (null generated by $n^a$)&with $w=\frac{1}{3}$&     \\
  \hline
\end{tabular}
  \caption{\small\sl Classification and examples of dynamical horizons. A specific dynamical horizon may be a union of various kinds of cases.}\label{CE}
\end{table}

Note that there is no restriction on the signature of the dynamical
horizon in our definition. In particular, if $\pounds_n\theta_l$
does not vanish, we can always choose a vector field $v^a=l^a-fn^a$
for some $f$ such that $v^a$ is tangential to the dynamical horizon.
It follows from $v_av^a=4f$ that the dynamical horizon is spacelike,
null, or timelike, depending on whether $f$ is positive, zero, or
negative, respectively. By virtue of $\pounds_v\theta_l=0$ and the
Raychaudhuri equation, we have on the dynamical horizon
\begin{equation}
f\pounds_n\theta_l=\pounds_l\theta_l=-\sigma^2-R_{ab}l^al^b,\label{Raychaudhuri}
\end{equation}
where $\sigma$ is the shear of $l^a$, and $R_{ab}$ is the Ricci
tensor. With the dominant energy condition satisfied by matter,
Eq.(\ref{Raychaudhuri}) implies that $\pounds_n\theta_l<0$ when the
dynamical horizon is spacelike, and $\pounds_n\theta_l>0$ when the
dynamical horizon is timelike. For the specific classification and
corresponding examples of dynamical horizons, please see Table
\ref{CE}.

On the other hand, our definition appears somewhat restrictive since
it rules out the degenerate cases in which $\theta_n=0$ or/and
$\pounds_l\theta_l=\pounds_n\theta_l=0$. Nonetheless, as
demonstrated in Table \ref{CE}, these degenerate cases rarely occur
in our interested dynamical spacetimes, thus our definition does not
lose its generality.

Now a covariant entropy bound conjecture on the general dynamical
horizon can be stated in a concise way: \emph{The entropy flux $S$
through the dynamical horizon between its apparent horizons of area
$A_d$ and $A_d'$ must satisfy $S\leq\frac{|A_d-A_d'|}{4}$ if the
dominant energy condition holds for matter.}

It is noteworthy that our general conjecture itself is manifestly
time reversal invariant. So its origin must be statistic rather than
thermodynamic, although it can be regarded as a reformulation of the
generalized second law of thermodynamics on the dynamical horizon in
some cases such as growing black holes and expanding
universes\cite{HZ1,HZ2}.
\section{Constraint $\Lambda$CDM cosmological models by the covariant entropy bound conjecture \label{C}}
In terms of the conformal time and comoving coordinates, the flat
FRW metric takes the form as
\begin{equation}
ds^2=a^2(\eta)[-d\eta^2+dr^2+r^2(dr^2+\sin^2\theta
d\phi^2)],
\end{equation}

Next, let us first compute the initial expansion of the
future-directed null congruences orthogonal to an arbitrary sphere
characterized by some value of $(\eta,r)$. Accordingly we obtain
\begin{equation}
\theta_{\pm}=\frac{\dot{a}}{a}\pm\frac{1}{r},\label{expansion}
\end{equation}
where the dot denotes the derivative with respect to $\eta$, and the
sign $+(-)$ represents the null congruence is directed at
larger(smaller) values of $r$. Thus the dynamical horizon here is
identified as
\begin{equation}
r_c(\eta)=\pm\frac{1}{h},\label{horizon}
\end{equation}
where $h\equiv\frac{\dot{a}}{a}$.

If as usual the matter content of FRW universes is assumed to be
described by the perfect fluid, with energy momentum tensor
\begin{eqnarray}
T_{ab}&=&a^2(\eta)\{\rho(\eta) (d\eta)_a(d\eta)_b+p(\eta)[(dr)_a
d(r)_b\nonumber\\
&&+r^2((d\theta)_a(d\theta)_b+\sin^2\theta(d\phi)_a(d\phi)_b)]\},
\end{eqnarray}
then by the Einstein equation with a positive cosmological constant
$\Lambda$, we have
\begin{eqnarray}
3h^2&=&8\pi\rho a^2+\Lambda a^2,\label{time}\\
-(h^2+2\dot{h})&=&8\pi p a^2-\Lambda a^2,\label{space}
\end{eqnarray}
From here, we can further obtain
\begin{eqnarray}
\dot{h}&=&-\frac{4\pi}{3}[(1+3w)\rho-\lambda] a^2,\label{derivative}\\
h^2-\dot{h}&=&4\pi (1+w)\rho a^2,\label{combination}
\end{eqnarray}
where $-1\leq w\leq1$ due to the dominant energy condition, and
$\lambda\equiv\frac{\Lambda}{4\pi}$.

To proceed, we further assume that the evolution of FRW universes is
adiabatical, which implies the conservation of the entropy current
associated with the matter, i.e., $\nabla_as^a$=0. Whence the
entropy current can be formulated as
\begin{equation}
s^a=\frac{s}{a^4}(\frac{\partial}{\partial\eta})^a,
\end{equation}
where $s$ is actually the ordinary comoving entropy density,
constant in space and time.

\begin{figure}[htb!]\small
\begin{center}
\includegraphics[clip=truth,width=0.8\columnwidth]{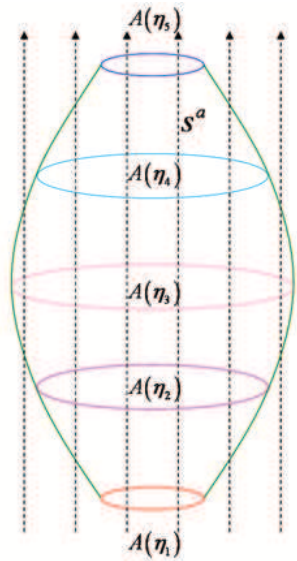}
\end{center}
\caption {\small \sl The dynamical horizon with the entropy current
flowing through it for $\Lambda$CDM cosmological models in the
conformal coordinates. When $\dot{r}_c\leq 0$, the entropy current
flows across the dynamical horizon from the interior region to the
exterior one while it flows from the exterior region to the interior
one for $\dot{r}_c\geq 0$.} \label{cdh}
\end{figure}

On the other hand, according to Eq.(\ref{horizon}), we have
\begin{equation}
\dot{r}_c=\mp\frac{\dot{h}}{h^2}.
\end{equation}
Note that at any moment the area of the dynamical horizon is give by
$A=4\pi a^2r_c^2$. Accordingly we can work out its time derivative,
i.e.,
\begin{equation}
\dot{A}=8\pi
a^2r_c^2(h+\frac{\dot{r}_c}{r_c})
=\frac{8\pi a^2(h^2-\dot{h})}{h^3}.
\end{equation}
Obviously, by Eq.(\ref{combination}), the increase or decrease of
area with time only depends on whether the universe is expanding or
contracting. In what follows we shall only focus on the expanding
universes, i.e., $h\geq 0$, where the corresponding area
monotonically increases with the evolution of time.

We shall now explore how our covariant entropy bound conjecture
provides an intriguing constraint on $\Lambda$CDM cosmological
models, where
$\rho=\frac{\rho_r^0a_0^4}{a^{4}}+\frac{\rho_m^0a_0^3}{a^{3}}$.
However, as demonstrated in Figure \ref{cdh}, it is noteworthy that
there is an obvious difference between $\dot{r}_c\leq 0$ and
$\dot{r}_c\geq 0$. Thus employing the conservation of the entropy
current and Gauss theorem, our conjecture can be equivalently
expressed as
\begin{equation}
\frac{\dot{A}}{4}+\dot{S}\geq 0
\end{equation}
for $\dot{r}_c\leq 0$($\dot{h}\geq 0$), and
\begin{equation}
\frac{\dot{A}}{4}-\dot{S}\geq 0
\end{equation}
for $\dot{r}_c\geq 0$($\dot{h}\leq 0$). Here $S$ denotes the entropy
flux through the interior region $r\leq r_c$, given by
$S=\frac{4\pi}{3} sr_c^3$, whereby we have
\begin{equation}
\dot{S}=4\pi sr_c^2\dot{r}_c=-\frac{4\pi
s\dot{h}}{h^4}.
\end{equation}
So by Eqs.(\ref{time}), (\ref{derivative}), and (\ref{combination}),
our conjecture gives
\begin{equation}
s\leq \frac{\sqrt{3(8\pi\rho a^2+\Lambda a^2)}(1+w)\rho
a^2}{2[(1+3w)\rho-\lambda]}\label{early}
\end{equation}
when $(1+3w)\rho>\lambda$, and
\begin{equation}
s\leq \frac{\sqrt{3(8\pi\rho a^2+\Lambda a^2)}(1+w)\rho
a^2}{2[\lambda-(1+3w)\rho]}\label{late}
\end{equation}
when $(1+3w)\rho<\lambda$, which apparently corresponds to the later
stages of expanding universes. In particular, to guarantee that the
bound (\ref{late}) holds at the very remote future, we thus obtain
the novel inequality (\ref{novel}) by setting the present scale
factor $a_0=1$.

For our own universe, as is well known, $\Lambda\sim 10^{-120}$ and
$\rho^0_m\sim\frac{1}{3}\Lambda$, so our conjecture follows that the
present entropy density should be less than $10^{-60}$, which is
satisfied with a wide safety margin, since the realistic entropy
density, dominated by super-massive black holes, is around of order
$10^{-75}$ today\cite{Frampton}. That is to say, our conjecture
supports the existence of our own universe as it should do. On the
other hand, if we take the present entropy density and dust energy
density as input data, our conjecture gives a novel upper bound on
the cosmological constant, i.e., $\Lambda<10^{-90}$, which obviously
much alleviates the cosmological constant problem why the
cosmological constant is so small in Planck units. Last but not
least, the presence of cosmological constant, albeit small, appears
to be in favor of the anthropic principle: To make our conjecture
satisfied, there should be the dust matter in our universe, which is
assumed to be a very basic condition for the creation of life.
Furthermore, due to the fact that the dust matter must be massive,
our conjecture seems to indicate a new close tie between our large
scale fiducial $\Lambda$CDM cosmological model and our small scale
standard model or beyond for
 elementary particles, namely, the origin of mass for the ordinary
matter and cold dark matter is determined to be intertwined with the
present cosmological constant somehow or others.
\section{Conclusion and discussion}
Advance in fundamental physics has often been driven by the
recognition of a new principle, a key insight to guide the search
towards a successful theory. In the ongoing search for a complete
theory of quantum gravity, the holographic principle stands out as
such a principle, which, in essence, relates geometric aspects of
spacetime to the number of quantum states of matter in a
surprisingly strong way, believed to be a law of physics that
captures one of the most crucial aspects of quantum gravity. As a
compelling pattern for the holographic principle, our covariant
entropy bound conjecture has been addressed for more general
dynamical horizons.

On the other hand, without knowledge of its microscopic makeup and
specific dynamics, the use of general principles to investigate a
system can be very rewarding. Thus we have also applied our proposed
covariant entropy bound conjecture to $\Lambda$CDM cosmological
models. As a result, it is shown that our conjecture implies a
remarkable upper bound $10^{-90}$ on the cosmological constant for
our own universe, which thus opens an alternative macroscopic
perspective to shed light on the longstanding cosmological constant
problem. In addition, our conjecture also indicates that there may
be a certain profound connection among the presence of the
cosmological constant, the origin of mass, and the anthropic
principle.

We conclude with an honest caveat. Although the results obtained so
far are particularly attractive as well as consistent with our
observational data, there remains a possibility that our starting
conjecture proves incorrect. It may be quite successful in many
respects only as a coincidence, but one should regard it as as a
warning, showing that our covariant entropy bound conjecture may
require a certain reformulation where it is violated, rather than as
a criterion or tool to constrain various models. Therefore it is
clear that both to provide more indirect or peripheral
justifications and to signify a deeper origin of our conjecture in
an underlying quantum theory of gravity are needed.
\section*{Acknowledgements}
Much gratitude should be given to Prof. Abhay Ashtekar for his
stimulating series of lectures at BNU, which directly induced us to
investigate the holographic implication of dynamical horizons in a
covariant way. HZ has also enjoyed valuable conversations with Prof.
Rafael Sorkin. Work by SH was supported by NSFC(nos.10235040 and
10421003). HZ was supported in part by the Government of China
through CSC(no.2007102530). This research was supported by Perimeter
Institute for Theoretical Physics. Research at Perimeter Institute
is supported by the Government of Canada through IC and by the
Province of Ontario through MRI.

\end{document}